\documentclass[francais]{gretsi}

\usepackage[utf8]{inputenc}

\usepackage[T1]{fontenc}

\usepackage{amsmath, amssymb, amsfonts}

\usepackage{enumitem}
\usepackage{caption,xcolor}
\begin{document}

\titre{Modèles de diffusion pour la reconstruction polarimétrique d'environnements circumstellaires}

\auteurs{
  \auteur{Quentin}{VILLEGAS}{quentin.villegas@oca.eu}{1}
  \auteur{Laurence}{DENNEULIN}{laurence.denneulin@epita.fr}{2,4}
  \auteur{Simon}{PRUNET}{simon.prunet@oca.eu}{1}
  \auteur{André}{FERRARI}{andre.ferrari@oca.eu}{1}\\
  \auteur{Nelly}{PUSTELNIK}{}{3}
  \auteur{Éric}{THI\'EBAUT}{}{4}
  \auteur{Julian}{TACHELLA}{}{3}
  \auteur{Maud}{LANGLOIS}{}{4}
}

\affils{
  \affil{1}{Université Côte d'Azur, CNRS, Observatoire de la Côte d'Azur, Nice, France}
  \affil{2}{Laboratoire de Recherche de l'EPITA, EPITA, 94270 Le Kremlin-Bicêtre, France}
  \affil{3}{Laboratoire de Physique, ENSL, CNRS UMR 5672, F-69342, Lyon, France}
 \affil{4}{Univ Lyon, Univ. Lyon 1, ENS de Lyon, CNRS, Centre de Recherche Astrophysique de Lyon, France},
}

\resume{Dans cet article, nous proposons une approche combinant modèles de diffusion et problème inverse pour la reconstruction d'images de disques circumstellaires. Notre méthode s'appuie sur le  modèle Rhapsodie pour l'imagerie polarimétrique, en substituant à son \textit{a priori} classique un modèle de diffusion entraîné sur des données synthétiques. Notre formulation intègre explicitement les fuites stellaires tout en traitant efficacement les données manquantes et le bruit élevé inhérents à l'imagerie polarimétrique à haut contraste. Les expériences montrent une amélioration significative par rapport aux méthodes conventionnelles dans le cadre de nos hypothèses, ouvrant de nouvelles perspectives pour l'étude des environnements circumstellaires.}

\abstract{In this paper, we propose an approach combining diffusion models and inverse problems for the reconstruction of circumstellar disk images. Our method builds upon the Rhapsodie framework for polarimetric imaging, substituting its classical prior with a diffusion model trained on synthetic data. Our formulation explicitly incorporates stellar leakage while efficiently handling missing data and high level noise inherent to high-contrast polarimetric imaging. Experiments show significant improvement over conventional methods within our framework of assumptions, opening new perspectives for studying circumstellar environments.}

\maketitle


\section{Introduction}
L'observation des disques circumstellaires est fondamentale pour comprendre la formation planétaire, mais demeure difficile en raison du contraste extrême avec leur étoile hôte, cette dernière étant typiquement 1 000 à 10 000 fois plus brillante. Malgré les instruments avancés comme SPHERE/IRDIS \cite{beuzit_sphere_2019}, les observations restent dégradées par divers facteurs instrumentaux et atmosphériques.

Rhapsodie \cite{denneulin_rhapsodie_2021} a récemment apporté des améliorations dans la reconstruction d'images circumstellaires en polarimétrie par une approche de problème inverse, permettant une meilleure gestion des données manquantes et du bruit. Par la suite, des techniques d'apprentissage profond par réseaux de neurones déroulés ont été proposées pour améliorer ces reconstructions \cite{chappon_linear_2024}. Bien qu'efficaces, ces méthodes nécessitent le calcul de gradients à travers toutes les itérations pendant l'entraînement, générant une empreinte mémoire importante, et doivent être réentraînées pour chaque nouvelle configuration instrumentale.

Notre travail propose une approche alternative basée sur les modèles de diffusion: plutôt que d'utiliser des méthodes classiques de régularisation, nous exploitons les modèles de diffusion probabilistes (DDPM) \cite{ho_denoising_2020}, capables d'apprendre efficacement la distribution statistique des images astrophysiques. Ces modèles transforment progressivement une distribution complexe en bruit gaussien puis apprennent à inverser ce processus. Ils ont démontré des performances remarquables en génération d'images et commencent à être appliqués aux problèmes inverses \cite{daras_survey_2024}.

Notre contribution principale est l'adaptation de l'algorithme Score-ALD (Annealed Langevin Dynamics) \cite{jalal_robust_2021} pour l'échantillonnage conditionnel à partir d'observations dégradées de disques circumstellaires. 
Notre modèle s'appuie sur une architecture de type U-Net inspirée de DDPM \cite{ho_denoising_2020}, combinant blocs résiduels et mécanismes d'auto-attention. Ce réseau est entraîné sur des données synthétiques générées via DDiT \cite{olofsson_challenge_2020}.
Soulignons qu'un avantage notable par rapport à \cite{chappon_linear_2024} réside dans l'apprentissage qui est ici réalisé indépendamment du modèle d'observation, donc avec une empreinte mémoire et un coût de calcul inférieurs. Les méthodes Plug-and-Play (PnP) \cite{zhang_plug-and-play_2021} offrent également une alternative intéressante, bien qu'elles requièrent un ajustement minutieux des hyperparamètres.

L'avantage majeur de notre approche est sa capacité à générer des échantillons de la distribution \textit{a posteriori}. En combinant l'échantillonnage conditionnel avec le modèle direct de Rhapsodie et une estimation de l'amplitude des fuites stellaires, la méthode proposée parvient à reconstruire efficacement les structures circumstellaires, même à partir de données fortement bruitées. Les résultats expérimentaux tendent à montrer une nette amélioration dans la récupération des détails morphologiques des disques, offrant ainsi de nouvelles perspectives pour l'étude détaillée des environnements circumstellaires.

\section{Formalisation du problème}

\subsection{Modèle d'observation en polarimétrie}

L'observation directe des environnements circumstellaires présente un défi majeur en raison du contraste élevé entre l'étoile et son disque d'accrétion. La lumière totale reçue $I$ combine plusieurs contributions : la lumière stellaire $I_{\text{star}}$ (non polarisée), et celle du disque $I_{\text{disk}}$ qui peut être décomposée en composante non polarisée $I_{\text{disk}}^{u}$ et polarisée $I_{\text{disk}}^{p}$ \cite{denneulin_rhapsodie_2021}: 
\begin{equation}
    I_\text{disk}= I_{\text{disk}}^{u} + I_{\text{disk}}^{p},\; I = I_\text{disk} + I_{\text{star}}
\end{equation}

L'objectif en imagerie polarimétrique est d'accéder aux paramètres $(I^u, I^p)$ où:
\begin{itemize}[nosep]
\item $I^u = I^u_{\text{star}} + I^u_{\text{disk}}$ représente l'intensité totale de la lumière non polarisée, combinant la contribution directe de l'étoile et celle du disque.
\item $I^p=I^p_{\text{disk}}$ correspond à l'intensité de la lumière polarisée, résultant de la diffusion de la lumière stellaire sur les poussières du disque. Ce paramètre est essentiel pour l'étude des processus de formation stellaire.
\end{itemize}

Les observations avec SPHERE-IRDIS exploitent la modulation de polarisation pour séparer les différentes composantes. Comme détaillé dans \cite{denneulin_rhapsodie_2021}, l'instrument utilise une lame demi-onde (HWP) aux angles $\alpha_\ell \in \{0^{\circ}, 22.5^{\circ}, 45^{\circ}, 77.5^{\circ}\}$ et deux analyseurs aux angles $\psi_j = \{0^{\circ}, 90^{\circ}\}$ pour produire différentes combinaisons de polarisation. La lumière détectée pour chaque configuration $(j,\ell)$ s'exprime par :
\begin{equation}
   I_{j,\ell}^{\text{det}} = \frac{1}{2}I^{u} + I^{p}\cos^2(\theta - 2\alpha_\ell - \psi_j)
\label{eq:eqdet}
\end{equation}
où $\theta$ est l'angle de polarisation. Cet angle dépend des propriétés de diffusion de la poussière dans le disque. Dans notre approche, nous adoptons une hypothèse fondamentale : pour une étoile ponctuelle et centrée (ce qui est le cas ici), et dans l'hypothèse de diffusion simple, l'angle de polarisation $\theta$ est connu car la polarisation est orthoradiale \cite{engler_detection_2018}. Cette propriété physique rend le modèle (\ref{eq:eqdet}) linéaire par rapport aux seuls paramètres inconnus qui sont l'intensité $I$ et $I^{p}$. Soulignons que contrairement à Rhapsodie, nous utiliserons donc les paramètres $(I, I^p, \theta)$ qui, contrairement aux paramètres de Stokes, ne vérifient pas de contraintes complexes.

Le modèle d'observation complet reliant les données mesurées $y_{j,\ell}$ à cette intensité détectée s'écrit :
\begin{equation}
   y_{j,\ell} = M_{j,\ell} T_{j,\ell} K I_{j,\ell}^{\text{det}} + \varepsilon_{j,\ell}
\label{eq:y_data_acquisition}
\end{equation}
où $K$ représente la convolution par la PSF (\textit{Point Spread Function}) instrumentale, $T_{j,\ell}$ les transformations géométriques, $M_{j,\ell}$ un masque binaire identifiant les pixels invalides qui représentent environ 10\% des données  et $\varepsilon_{j,\ell}$ est le bruit d'observation supposé gaussien avec une variance qui dépend de l'intensité du pixel (bruit hétéroscédastique) et de la configuration d'acquisition.

\subsection{Formulation du problème inverse}
\label{sec:formulation_pb_inverse}

Nous supposerons dans la suite qu'il est possible de modéliser de façon explicite  la contamination stellaire:
\begin{equation}
   I = I_{\text{disk}} + \lambda \cdot I_{\text{star}}
\end{equation}
où $\lambda$ est le coefficient de fuite stellaire supposé inconnu et $I_{\text{star}}$ l'intensité calibrée de l'étoile coronographiée. Cette paramétrisation permet de séparer efficacement les composantes, crucial pour l'étude précise des disques faiblement lumineux.

Notre modèle est donc caractérisé par les paramètres $(I_{\text{disk}}, I^{p}, \lambda)$. Nous utilisons le modèle direct développé dans Rhapsodie \cite{denneulin_rhapsodie_2021} qui, en regroupant tous les opérateurs de \eqref{eq:y_data_acquisition} en un unique opérateur $\mathsf{A}$, peut s'écrire sous la forme condensée :
\begin{equation}    
  y = \mathsf{A} x + \lambda   \mathsf{B} s + \varepsilon
  \label{eq:foward_model}
\end{equation}
où $x = (I_{\text{disk}}, I^p)$ est l'image du disque seul, $s=(I_{\text{star}},0)$ représente les contaminations stellaires incluant la PSF coronographique, $y$ les données observées, $\mathsf{A}$ l'opérateur de dégradation linéaire (calculant $I_{\text{disk}}^u=I_{\text{disk}}-I^p$ avant d'appliquer les convolutions, transformations géométriques et analyseurs), $\mathsf{B}$ identique à $\mathsf{A}$ dépourvu de convolution ($K=\mathrm{I}$) et $\varepsilon$ un bruit gaussien indépendant de moyenne nulle décrit dans \eqref{eq:y_data_acquisition}. La matrice de précision de $\varepsilon$, notée $W$, est une matrice diagonale avec des entrées nulles pour les pixels invalides (morts, saturés ou masqués). Avec ces notations, on obtient:
\begin{equation}
\nabla_{x} \log p(y|x, \lambda) = \mathsf{A}^\dagger\mathsf{W}(\mathsf{A}x+\lambda \mathsf{B} s-y)
\label{eq:grad}
\end{equation}
où  les données corrompues sont naturellement écartées.

Afin de reproduire les propriétés d'un bruit de photons à fort flux lors de la simulation des données, la variance des pixels valides est la somme de la variance du bruit de lecture et d'un terme proportionnel à l'intensité du pixel. 


\section{Algorithme de reconstruction}

\subsection{Processus de diffusion} 

Le premier élément de la reconstruction est l'estimation d'un \textit{a priori} sur les images polarisées et non polarisées du disque $x=(I_\text{disk},I^p)$. Celui-ci est obtenu par un modèle de diffusion. Ces modèles \cite{ho_denoising_2020} considèrent une transformation progressive d'une distribution complexe en bruit gaussien. Le processus direct s'écrit: 
\begin{equation}
q(x_t|x_{t-1}) = \mathcal{N}(x_t; \sqrt{1-\beta_t}x_{t-1}; \beta_t I)
\end{equation}
où $\{x_t\}_{t=1}^T$ représente des versions de plus en plus bruitées de l'image $x_0$, et $\{\beta_t\}_{t=1}^T$ est une séquence croissante de variances . 

Ce processus markovien se formule aussi directement entre $x_0$ et $x_t$:
\begin{equation}
x_t = \sqrt{\bar{\alpha}_t}x_0 + \sqrt{1-\bar{\alpha}_t}\epsilon
\end{equation}
où $\epsilon \sim \mathcal{N}(0, I)$ et $\bar{\alpha}_t = \prod_{i=1}^t (1-\beta_i)$. L'objectif  est d'inverser ce processus pour reconstruire une image à partir du bruit gaussien $x_T$.

Pour réaliser cette inversion, l'approche DDPM \cite{ho_denoising_2020} s'appuie sur l'apprentissage d'un réseau $\epsilon_\theta(x_t, t)$ capable de prédire le bruit $\epsilon$ ayant transformé l'image $x_0$ en sa version bruitée $x_t$. L'objectif d'optimisation s'écrit:
\begin{equation}
\mathcal{L}_{\text{simple}}(\theta) = \mathbb{E}_{t, x_0, \epsilon} \left[ \|\epsilon-\epsilon_\theta(\sqrt{\bar{\alpha}_t}x_0 + \sqrt{1-\bar{\alpha}_t}\epsilon, t) \|^2 \right]
\label{eq:l_simple}
\end{equation}
Une propriété fondamentale établie par \cite{dhariwal_diffusion_2021} montre que cette formulation est équivalente à l'apprentissage du score de la distribution $\nabla_{x_t} \log p_t(x_t)$. En effet, $x_0|x_t$ étant Gaussien, le score du noyau de transition de $x_0$ à $x_t$ s'exprime par :
\begin{equation}
\nabla_{x_t} \log p(x_t|x_0) = -\frac{x_t - \sqrt{\bar{\alpha}_t}x_0}{1-\bar{\alpha}_t} = -\frac{\epsilon}{\sqrt{1-\bar{\alpha}_t}}
\end{equation}
Cette relation implique que minimiser $\mathcal{L}_{\text{simple}}$ \eqref{eq:l_simple} revient à entraîner le réseau $\epsilon_\theta$ à approximer une version mise à l'échelle du score du noyau de transition. De plus, \cite{song_score-based_2021} démontre que l'espérance de ce score correspond à celui de la distribution $p_t(x_t)$ par la relation :
\begin{equation}
\nabla_{x_t} \log p_t(x_t) = \mathbb{E}_{p(x_0|x_t)}\left[\nabla_{x_t} \log p(x_t|x_0)\right]
\end{equation}
Ainsi, la procédure d'entraînement standard des DDPM, qui consiste à prédire le bruit $\epsilon$, est une méthode efficace pour apprendre une version mise à l'échelle du score. Une fois le réseau entraîné, nous pouvons estimer le score de la distribution marginale en utilisant la relation:
\begin{equation}
\nabla_{x_t} \log p_t(x_t) \simeq -\frac{1}{\sqrt{1-\bar{\alpha}_t}}\epsilon_\theta(x_t,t)
\end{equation}
Cette formulation établit un lien direct entre les modèles de diffusion et l'estimation du score, unifiant ainsi deux approches initialement développées séparément.

\subsection{Méthode proposée pour la reconstruction}

Pour résoudre notre problème inverse, nous adaptons l'approche Score-ALD (Annealed Langevin Dynamics) \cite{jalal_robust_2021}
pour reconstruire le disque $x$. Notre algorithme combine l'échantillonnage de Langevin avec le score du modèle de diffusion conditionné par les observations. À partir du théorème de Bayes, le score du modèle conditionnel à $\lambda$, $\nabla_{x_t} \log p(x_t|y,\lambda)$ est approché par :
\begin{equation*}
\nabla_{x_t} \log p(x_t)  - \mathsf{A}^\dagger\mathsf{W}(\mathsf{I} + \gamma_t^2\mathsf{W})^{-1}(\mathsf{A}x_t+\lambda \mathsf{B}s-y).
\end{equation*}
Le premier terme correspond au score du modèle de diffusion, tandis que le second représente la fidélité aux données observées \eqref{eq:grad} où $\gamma_t$ est le terme de recuit simulé, décroissant avec $t$ et $\mathsf{W}$ la matrice de poids décrite dans la section \ref{sec:formulation_pb_inverse}. 

Pour $\lambda$ fixé, l'algorithme Score ALD peut ainsi être appliqué pour générer des échantillons à partir de la distribution postérieure $p(x|y,\lambda)$. Afin d'estimer conjointement à $x$ le coefficient de fuite 
stéllaire $\lambda$, nous proposons de mettre à jour à chaque itération son estimation en utilisant :
$$
\lambda_{t-1} = \arg\max_\lambda p(\lambda|x_{t-1},y) = \frac{s^\dagger \mathsf{B}^\dagger\mathsf{W}(y-\mathsf{A}x_{t-1})}{\|\mathsf{B}s\|^2_\mathsf{W}}
$$
avec une distribution \emph{a priori} $p(\lambda)$ uniforme. La procédure d'estimation complète est présenté dans Algorithme \ref{alg:score_ald} où  $T$, $\{\beta_t\}_{t=1}^{T}$ et $\{\bar{\alpha}_t\}_{t=1}^{T}$ sont fixés lors de l'entraînement de $\epsilon_\theta$. Les hyperparamètres pour l'inférence sont les $\{\gamma_t\}_{t=1}^{T}$.

\begin{algorithm}[t]
Initialiser $x_T \sim \mathcal{N}(0, I)$ \\
\Pour{$t=T, T-1, \ldots, 1$}{
  $\zeta_t \sim \mathcal{N}(0 ; I)$ \\
  $x_{t-1} \leftarrow x_t +\sqrt{2 \beta_t} \zeta_t - \beta_t \left(\frac{\epsilon_\theta(x_t, t)}{\sqrt{1-\bar{\alpha}_t}} \right.$+ \\ 
  \hfill $\left. \mathsf{A}^\dagger\mathsf{W}(\mathsf{I} + \gamma_t^2\mathsf{W})^{-1}(\mathsf{A}x_t+\lambda  \mathsf{B} s-y) \right) $ \\
  $\lambda_{t-1}\leftarrow\frac{s^\dagger \mathsf{B}^\dagger\mathsf{W}(y-\mathsf{A}x_{t-1})}{\|\mathsf{B}s\|^2_\mathsf{W}}$
}
\caption{Algorithme Score-ALD proposé.}
\label{alg:score_ald}
\end{algorithm}

L'hypothèse $s$ connu, permet ainsi de raffiner progressivement l'estimation de la contamination stellaire. Bien que cette hypothèse constitue une simplification, elle offre un premier pas vers une séparation des composantes stellaire et circumstellaire.


\section{Expériences et résultats}

\subsection{Jeux de données et entraînement}
Pour évaluer la performance de notre modèle de diffusion, nous avons constitué un ensemble de données synthétiques avec l'outil DDiT (Debris DIsks Tool) \cite{olofsson_challenge_2020}. Ce logiciel permet de simuler l'intensité polarisée des disques de débris dans les environnements circumstellaires.

DDiT génère des données réalistes en simulant la géométrie 3D des disques et la diffusion de la lumière. Cependant, ces simulations reposent sur des hypothèses simplificatrices : la diffusion simple en milieu optiquement mince et une distribution de poussière lisse, ignorant les surdensités locales et diffusions multiples.

Notre jeu de données a été généré en faisant varier plusieurs paramètres physiques comme le demi-grand axe du disque, l'inclinaison, l'excentricité ou l'angle d'ouverture. Cette paramétrisation nous a permis d'obtenir les intensités $(I^u_{disk}, I^p)$. Pour reproduire des conditions d'observation réalistes, nous avons ajouté une composante stellaire $I_{star}$ issue de données coronographiques à haut contraste provenant de l'instrument SPHERE \cite{beuzit_sphere_2019}, obtenant ainsi l'intensité non polarisée totale $I^u=I^u_{disk}+I_{star}$.
20000 images de dimension $128 \times 128$ pixels ont été utilisées avec différents niveaux de bruit, correspondant aux valeurs $t$ du processus de diffusion, pour entraîner notre modèle suivant l'équation \eqref{eq:l_simple}. Le réseau $\epsilon_\theta$ basé sur un UNet totalise 80,4 millions de paramètres et son entraînement a été réalisé en environ 40 heures sur un GPU NVIDIA A100.

\subsection{Évaluation des performances}
Nous avons évalué notre méthode sur des données synthétiques représentatives des observations réelles. Pour ce faire, nous avons généré un jeu de test avec DDiT comprenant des disques n'appartenant pas à l'ensemble d'entraînement. 

\begin{figure}
\begin{center}
\includegraphics[width=\columnwidth]{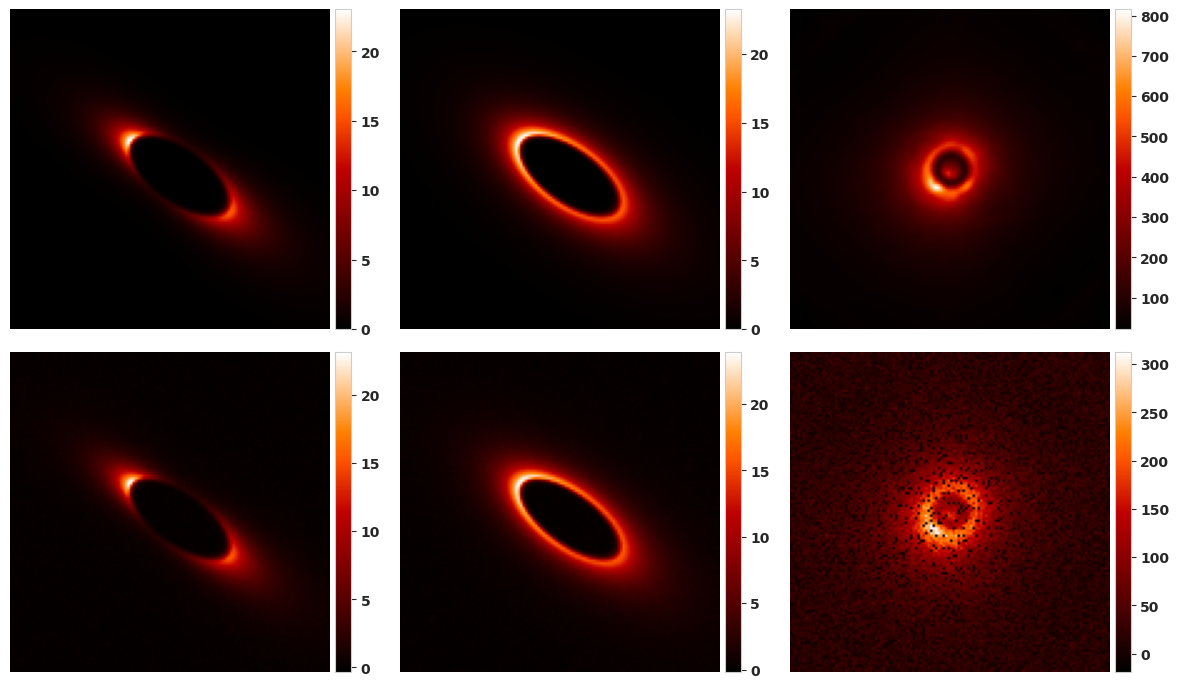}
\end{center}
\vspace{-8pt}
\caption{Ligne 1 : $I^p$, $I_{\text{disk}}$ et $I_{\text{disk}}+I_{\text{star}}$. Ligne 2  : reconstruction $I^p$, $I_{\text{disk}}$ et une des 4 mesures \eqref{eq:y_data_acquisition} (ici $y_{0,0}$). \label{fig:recon}}
\end{figure}

Notre cas de test principal représente un scénario difficile : un disque circumstellaire $35$ fois moins brillant que les fuites stellaires, observé avec seulement 4 mesures (une acquisition par position de lame demi-onde). Cette configuration correspond à des conditions d'observations à haut contraste, où le rapport rapport signal-sur-bruit (SNR) est faible. En effet  le SNR  $I_{\text{disk}}+I_{\text{star}}$ est de 14 dB alors que celui de $I_{\text{disk}}$ est de -16 dB. La figure \ref{fig:recon} montre un exemple de reconstruction pour la méthode proposée.

La table \ref{tab:resultats} présente une comparaison quantitative de notre méthode avec la méthode Rhapsodie \cite{denneulin_rhapsodie_2021}, utilisant une régularisation à préservation de contours. Les résultats indiquent une nette amélioration apportée par notre approche, avec un gain en PSNR d'environ 14 dB par rapport à Rhapsodie. Les métriques d'erreur MAE et RMSE montrent également une réduction substantielle, présentant des valeurs approximativement six fois plus faibles que celles obtenues avec Rhapsodie.

\begin{table}[t]

\begin{center}
\begin{tabular}{lccc}
\toprule
\textbf{Méthode} & \textbf{PSNR (dB)} & \textbf{MAE} & \textbf{RMSE} \\
\midrule
Rhapsodie & 24,74 & 0,0358 & 0,0574 \\
Méthode proposée & \textbf{39,28} & \textbf{0,0060} & \textbf{0,0108} \\
\bottomrule
\end{tabular}
\end{center}
\caption{\label{tab:resultats}Comparaison des performances moyennes de reconstruction calculées en utilisant 100 simulations.}
\end{table}

L'amélioration des performances peut s'expliquer par plusieurs facteurs : d'une part, la capacité de notre modèle de diffusion à capturer efficacement la distribution \textit{a priori} des disques circumstellaires, et d'autre part, notre stratégie d'estimation du coefficient de fuite stellaire qui raffine progressivement la séparation entre l'étoile et son environnement. Un élément important est aussi que Rhapsodie ne tient pas compte du fait que l'angle de polarisation est connu et va l'estimer.

Ces résultats doivent être nuancés par certaines limitations importantes. Notre modèle repose sur des hypothèses simplificatrices concernant les fuites stellaires, supposant une distribution connue pilotée uniquement par le coefficient $\lambda$. De plus, le modèle de diffusion a été entraîné exclusivement sur des images de disques sans contamination stellaire, ce qui explique sa difficulté potentielle à gérer des scénarios réels plus complexes. 
Une solution à ce problème consiste à utiliser un \textit{a priori} Gaussien pour $s$ basé sur des covariances spatiales estimées localement, comme proposé par \cite{flasseur_rexpaco_2021}.


\section{Conclusions et perspectives}
Nous avons présenté une nouvelle approche pour la reconstruction d'images de disques circumstellaires basée sur les modèles de diffusion probabilistes. Alors que des méthodes récentes ont exploré l'utilisation d'architectures déroulées pour ce type de problèmes \cite{chappon_linear_2024}, notre approche exploite la capacité des modèles de diffusion à capturer des distributions complexes pour régulariser le problème inverse.

Les résultats expérimentaux démontrent la pertinence de notre approche, tant en termes de métriques quantitatives que de qualité visuelle. Particulièrement, notre méthode s'avère très performante pour reconstruire des structures fines et des régions de faible intensité, cruciales pour l'analyse astrophysique.

Parmi les perspectives, d'autres méthodes plus sophistiquées \cite{daras_survey_2024} pourraient remplacer notre implémentation actuelle de Score-ALD. De même, l'adoption d'architectures plus récentes comme les modèles de diffusion latents \cite{song_solving_2024} améliorerait potentiellement la qualité des reconstructions tout en réduisant le temps de calcul. L'intégration de connaissances physiques spécifiques dans le processus de diffusion est aussi une perpsective importante.
Nous envisageons d'utiliser des modèles probabilistes de fuites stellaires et de les intégrer dans le processus de diffusion. Ces améliorations renforceraient la robustesse de notre approche dans le cadre d'observations réelles et permettraient d'exploiter pleinement le potentiel des modèles de diffusion pour l'imagerie astronomique à haut contraste.

\paragraph{Remerciements :} Ce travail a bénéficié de deux aides de l'État gérées par l'Agence Nationale de la Recherche au titre de France 2030 du PEPR ORIGINS pour le projet ciblé AIDA portant la référence ``ANR-22-EXOR-0016'' et au titre de l'ANR DDISK portant la référence ``ANR-21-CE31-0015''. \\
Ce travail a également bénéficié d’un accès aux ressources de calcul en IA et de stockage à l'IDRIS au travers de l'allocation de ressources 20XX-AD010415910 attribuée par GENCI sur la partition V100/A100 du calculateur Jean Zay.

\fontsize{9}{10}\selectfont
\bibliography{references}


\end{document}